# *Green steel at its crossroads: hybrid hydrogen-based reduction of iron ores*

Isnaldi R. Souza Filho[1,*], Hauke Springer[1,2], Yan Ma[1], Ankita Mahajan[1], Cauê C. da Silva[1], Michael Kulse[1], Dierk Raabe[1]

[1] *Max-Planck-Institut für Eisenforschung, Max-Planck-Str. 1, 40237 Düsseldorf, Germany*
[2] *Institut für Bildsame Formgebung, RWTH Aachen University, Intzestr. 10, 52072 Aachen, Germany*

[*] *corresponding author: i.souza@mpie.de*

**Abstract**

Iron- and steelmaking cause ~7% of the global $CO_2$ emissions, due to the use of carbon for the reduction of iron ores. Replacing carbon by hydrogen as the reductant offers a pathway to massively reduce these emissions. However, the production of hydrogen using renewable energy will remain as one of the bottlenecks at least during the next two decades, because making the gigantic annual crude steel production of 1.8 billion tons sustainable requires a minimum stoichiometric amount of ~97 million tons of green hydrogen per year. Another fundamental aspect to make ironmaking sector more sustainable lies in an optimal utilization of green hydrogen and energy, thus reducing efforts for costly in-process hydrogen recycling. We therefore demonstrate here how the efficiency in hydrogen and energy consumption during iron ore reduction can be dramatically improved by the knowledge-based combination of two technologies: partially reducing the ore at low temperature via solid-state direct reduction (DR) to a kinetically defined degree, and subsequently melting and completely transforming it to iron under a reducing plasma (i.e. via hydrogen plasma reduction, HPR). Results suggest that an optimal transition point between these two technologies occurs where their efficiency in hydrogen utilization is equal. We found that the reduction of hematite through magnetite into wüstite via DR is clean and efficient, but it gets sluggish and inefficient when iron forms at the outermost layers of the iron ore pellets. Conversely, HPR starts violent and unstable with arc delocalization, but proceeds smoothly and efficiently when processing semi-reduced oxides. We performed hybrid reduction experiments by partially reducing hematite pellets via DR at 700°C to 38% global reduction (using a standard thermogravimetry system) and subsequently transferring them to HPR, conducted with a lean gas mixture of Ar-10%$H_2$ in an arc-melting furnace, to achieve fully conversion into liquid iron. This hybrid approach allows to exploit the specific characteristics and kinetically favourable regimes of both technologies, while simultaneously showing the potential to keep the consumption of energy and hydrogen low and improve both, process stability and furnace longevity by limiting its overexposure to plasma radiation.

*Keywords:* green steel; hybrid hydrogen-based reduction of iron ores; efficiency in $H_2$ consumption; process stability.

# 1. Introduction

The urgent mission for decarbonisation has put the steel industry at the crossroads of a critical transition from fossil-based to hydrogen-based iron production [1–3]. This is because, currently, more than 70% of the global iron production is conducted through the blast furnace (BF) and basic oxygen furnace (BOF) integrated route, which generates the staggering amount of approximately 1.9 tons $CO_2$ per each ton of produced steel due to the use of carbon-carrying substances as reducing agents. Such a number makes the iron- and steelmaking responsible for about 7% of all $CO_2$ emissions on the planet, yet with a projection for a further massive increase for the coming decades [4,5]. Therefore, replacing carbon by green hydrogen (i.e. produced via renewable energy sources) as the reducing agent offers a promising pathway for cleaner and sustainable iron production, as the associated by-product is water rather than $CO_2$.

From the plethora of possible technological approaches to produce iron using hydrogen as a reducing agent, three have been identified as most viable for widespread industrial application: one is the direct injection of hydrogen into existing blast furnaces, in addition to coke (Figure 1a) [6–8]; the second one is the direct reduction (DR) conducted in shaft furnaces or fluidized bed reactors, where the solid iron ore is exposed to hydrogen-containing gases (Figure 1b) [9–11]; the third one is the hydrogen plasma-based reduction (HPR), in which the ores get melted and reduced simultaneously (Figure 1c) [12–14]. The first option has the highest technology-readiness level; however, it still requires coke as energy supply and porous solid support to permit gas percolation, and thus can only serve as a short-term transition technology of limited effectiveness. DR performed with reformed natural gases is a well-established route (e.g. the MIDREX process) , but operation with molecular hydrogen has been proven to be feasible as well [9,10,15]. Independent on the reducing agent used, though, DR provides solid sponge iron, which has to be subsequently melted, typically by conventional plasma in an electric arc furnace (EAF) to provide the liquid iron for alloy adjustment and casting [16], Figure 1 (d). The HPR process on the other hand is much more thermodynamically efficient as it allows for simultaneous reduction and melting of the ores as well as for the addition of scrap into the same aggregate [12,17,18], but it is still in its exploratory stage. However, many technical aspects required in reducing hydrogen plasma reactors are already standard technology in existing EAFs. For example, the operational and controlling conditions of electric/plasma arcs in EAFs are standard industrial practices as well as magnetics stirring of the molten baths [19]. When operated under a DC current mode, EAFs can be set with one single electrode, namely cathode, enabling the bottom of the furnace to serve as the anode, i.e. similar to the setup employed in HPR where the plasma arc is ignited between the tip of the electrode and the material to be processed (Figure 1c). Such configuration enables simultaneous melting and stirring of the charged material, as it exploits the heat from the DC current passing through the melt and the associated produced magnetic fields [19]. Electrode and crucible design, protective slags formation for lining preservation, and water-based cooling systems are transferable knowledge from industrial EAFs to hydrogen plasma reactors [20]. Thus, it is conceivable that slight adjustments of existing EAF technology for such hydrogen-based plasma-reduction purposes would require only comparatively modest investments, a key aspect for decision making in this industry at its current crossroads.

One essential task for making steel production not only sustainable but also commercially viable lies in the most efficient use of the costly reducing agent (green) hydrogen [21–23]. This is because of its direct link to the total energy balance which also includes the associated efforts for its in-process recycling and reutilization, as non-consumed $H_2$, steaming from the outflux streams, can be reintroduced back to the processes provided it is properly purified [15,24,25]. Driven by this goal, we monitored the reduction kinetics of iron ores processed under individual DR and HPR routes along with their corresponding efficiency in hydrogen utilization. We found that an optimal transitioning point between these two technologies occurs at the reduction stage where their efficiency in hydrogen utilization has a cross-over point. These findings suggest that the goal for an efficient usage of hydrogen in sustainable iron production (with minimum quantitates being sent to recycling) can be reached when combining DR and HPR into a novel hybrid process, where iron oxides partially reduced via a DR process to a kinetically defined degree is fed into a reducing plasma reactor to reach full conversion into liquid iron (Figure 1d).

Our findings demonstrate that this hybrid process not only occurs without compromising the total process time (as the iron sponge coming from DR must be subsequently melted anyway) but also keeps both, the usage of hydrogen and consumption of energy low, by exploiting the process efficiency limits of both DR and HPR. The reason for this synergy lies in the underlying physics of the reduction kinetics. The first reduction steps in hydrogen-based DR, from hematite through magnetite into wüstite, are very fast [26–29]. In contrast, the final reduction of wüstite into iron is very slow, endothermic, and requires a higher chemical potential of hydrogen. The thermodynamic requirements for wüstite reduction thus impose a utilization limit for hydrogen, causing additional efforts for its in-process recycling [30]. This means that DR starts efficiently (reducing hematite to wüstite), but gets slow and inefficient in the last stages of the reduction into iron [27,30,31]. The hydrogen-containing plasma on the other hand is violent and potentially unstable at the beginning of the reduction process, but becomes much smoother when the ore has started to liquefy. Thus, under well-controlled arc conditions, HPR can proceed more efficiently with using semi-reduced oxides. Also, limiting the exposure of the furnace to a reducing plasma substantially enhances its lifespan, particularly regarding refractory linings and electrodes [20,32].

Scientific results reported here help identify and understand the most kinetically and energetically favourable regimes of individual lab-scale hydrogen-based DR and HPR processes. A comprehensive evaluation of the associated underlying mechanisms that drive the reduction reactions yet enables the combination of these regimes into a hybrid and efficient strategy for iron production with low impurity contents. With this work, we aim to open a new perspective for decision making steps in green steel production towards the optimization of the main boundary conditions in this field. They include a reflected green hydrogen exploitation, permitting also its optimized recycling, as well as a minimum total energy consumption, and robust and stable processing conditions (e.g. plasma arc stability) at minimum investment and maintenance costs (e.g. by limiting inner furnace linings to overexposure to plasma). Otherwise the risk exists that non-sustainable hydrogen and inefficient energy usage will sweep the market, shifting greenhouse gas emissions simply from one industry to another, without enhancing industry's overall sustainability.

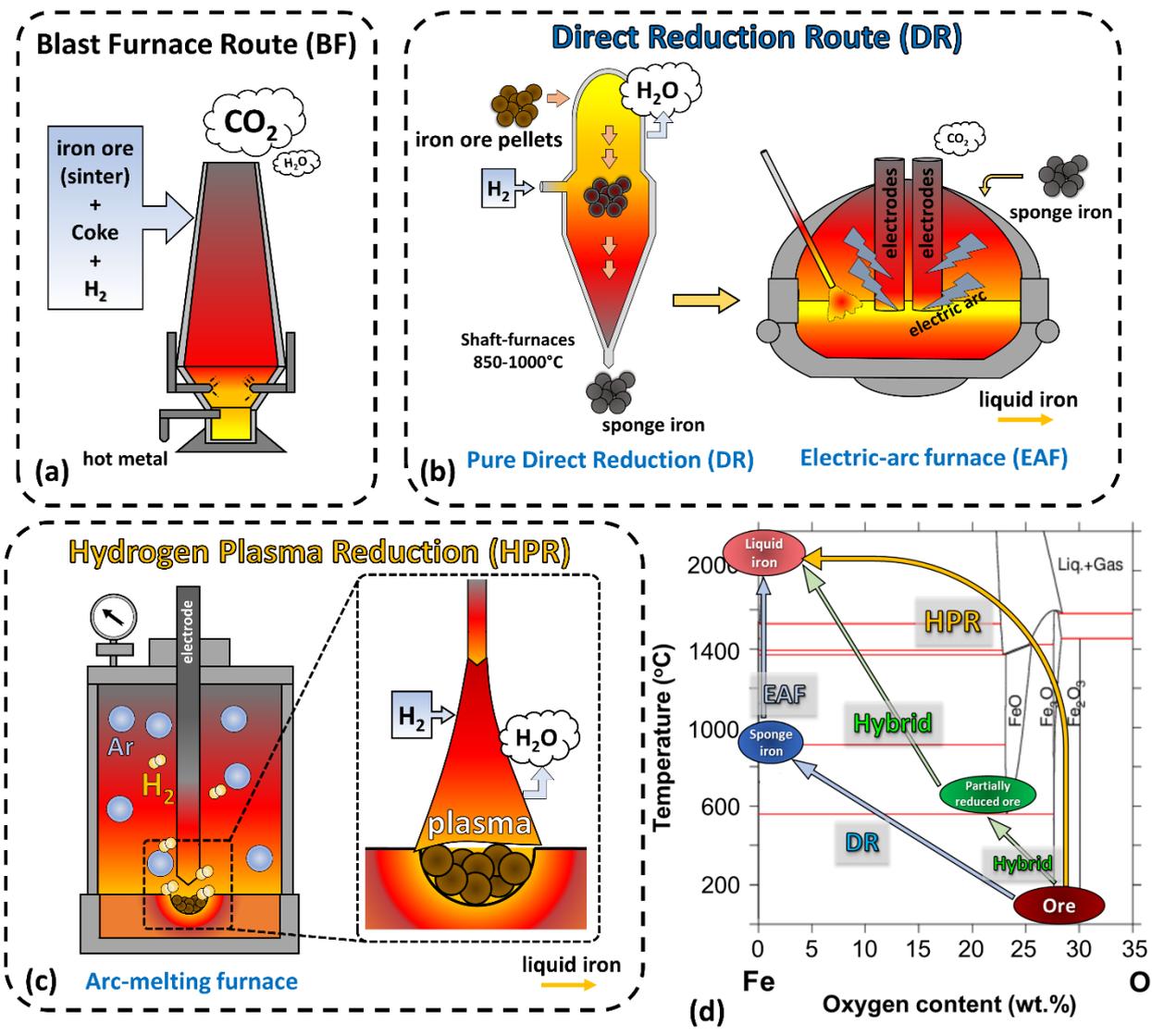

**Figure 1.** (**a**) Schematic representation of a blast furnace (BF) partially charged with $H_2$ in addition to coke for iron oxide reduction. (**b**) Schematic representation of the direct reduction route, in which iron ore pellets are converted into sponge iron via direct exposure to molecular hydrogen in a shaft furnace. The produced sponge iron is melted in a conventional electric arc furnace (EAF). (**c**) Schematic representation of the hydrogen plasma reduction of iron ores, conducted in an arc melting furnace, partially flooded with hydrogen. (**d**) Iron production routes projected onto a Fe-O phase diagram, highlighting DR, HPR, and the hybrid route.

## 2. Experimental

### 2.1 Material

In this work, commercial hematite in the form of both, pellets (average diameter of 8 mm) and irregular pieces (average size of 1.5 mm) were used. The chemical composition of the pellets is 28.2 wt.% O, 0.52 wt.% Si, 0.11 wt.% Mn, 0.18 wt.% Al, 0.50 wt.% Ca, 0.37 wt.% Mg, 0.11 wt.% Ti and traces of P, S, Na, K (with Fe as balance). The chemical composition of the irregular pieces is 29.4 wt.% O, 0.10 wt.% Si, 0.01 wt.% Mn, 0.08 wt.% Al, 0.13 wt.% S and traces of Ca, Mg, Ti (with Fe as balance). The content of metals was evaluated via inductively coupled plasma optical emission spectrometry (ICP-OES). The oxygen content was evaluated via reduction fusion under a helium atmosphere and subsequent infrared absorption spectroscopy. Sulphur was evaluated via combustion followed by infrared absorption spectroscopy.

### 2.2 Solid-state direct reduction

Hematite pellets were isothermally reduced into sponge iron in the static bed of a standard thermogravimetry apparatus at temperatures of 700 and 900°C under an $H_2$ flow of 0.5 L/min [33]. The static bed is connected to a thermal balance and encapsulated by a quartz glass tube, which is surrounded by a cylindrical-shaped infrared furnace. Figure 2 (a) shows a schematic representation of the thermogravimetry system in which the furnace is drawn in its open position. Before annealing, a constant $H_2$ flow of 0.5 L/min was set to percolate at room temperature through the quartz glass tube containing the material to be processed. This procedure was kept for 1 h in order to remove any traces of gaseous substances (especially oxygen and water). Subsequently to this cleaning procedure, isothermal annealing was conducted for 2 h at 700 and 900°C respectively. The changes in weight of the pellets were continuously tracked over the course of the reduction processes by the thermal balance. These values were further used to calculate the reduction degree as well the amount of $H_2$ effectively consumed, as reported in Section S.1 in the Supplementary Material. The pellets were cooled down to room temperature inside the furnace. Further details of this procedure can be found in a preceding study [30].

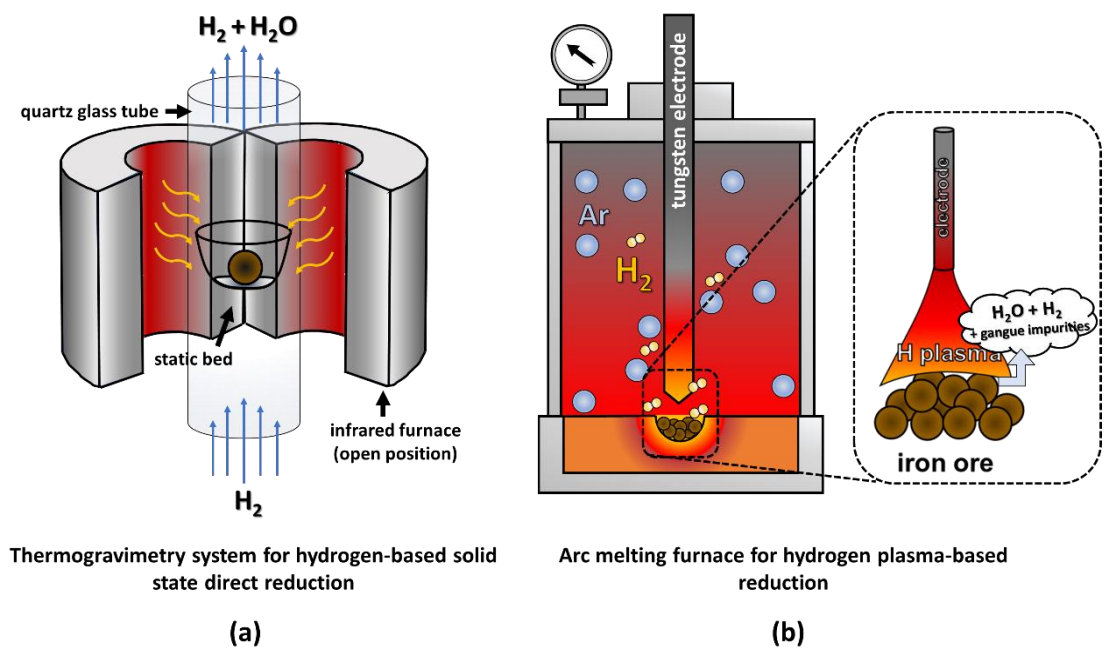

**Figure 2. (a)** Schematic representation of the thermogravimetry apparatus used for hydrogen-based solid-state direct reduction of iron ores [33]. The output stream is mostly composed of non-consumed $H_2$ and the by-product water vapour **(b)** schematic representation of the arc melting furnace equipped with a tungsten electrode and flooded with an Ar-10%$H_2$ gas mixture used for hydrogen plasma-based reduction of iron ores. The output stream is a complex mixture of non-consumed $H_2$, water vapour and gangue impurities evaporated from the ore [12].

## 2.3 Hydrogen-plasma reduction of molten iron ore

Hematite pieces with an average weight of 15 g were introduced in an arc-melting furnace equipped with a 6 mm-diameter tungsten electrode, Figure 2 (b). The furnace chamber (18 L) was charged with an Ar-10%$H_2$ mixture (total pressure of 900 mbar). Simultaneous melting and reduction of the samples were achieved by igniting an arc plasma between the tip of the electrode and the input iron oxide at a voltage of 44 V and a current of 800 A (Figure 2b). To achieve different intermediate levels of reduction, hematite pieces were submitted to a series of melting, solidification and remelting cycles up to 15 times. The corresponding plasma exposure was 1 min for each cycle where complete melting was achieved with only 6 s of exposure to hydrogen plasma. Samples were reduced during 1, 2, 5, 10, and 15 min. Proper reducing atmosphere conditions to maintain the progressing reaction were ensured by replenishing the furnace chamber with fresh Ar-10%$H_2$ after the completion of every single melting and solidification cycle. Further details of the experimental apparatus and reducing conditions are reported in a preceding study [12].

The solidified samples were hammered in order to separate the mm-sized iron portion from the remaining unreduced oxide one. The latter was further powdered and subjected to X-ray diffraction (XRD) measurements using a diffractometer D8 Advance A25-X1 (cobalt K$\alpha$ X-ray source), operated at 35 kV and 40 mA. The weight fraction of the phases contained in the powder was determined via Rietveld refinement. The weight fractions of the iron oxide variants (mainly wüstite) were reported in a preceding study [12]. Here, these values were used to calculate the corresponding amount of oxygen in all samples.

To evaluate the thermal decomposition of hematite and sort out the corresponding amount of oxygen that actually reacts with hydrogen and the one that is evaporated due to vapour pressure, 15 g of hematite pieces were processed in the arc-melting furnace under a pure argon atmosphere using the same process conditions described above (i.e. 44 V, 800 A, total pressure of 900 mbar, up to 15 min process). The furnace chamber was also replenished with fresh argon after the completion of every single melting cycle. Samples were also hammered into powder and characterized via XRD measurements according to the above-mentioned protocol. The corresponding amounts of consumed hydrogen were determined via mass balance calculation, as reported in Section S.1 in the Supplementary Material.

## 2.4 Hybrid reduction

Hematite pellets were partially reduced via DR (Figure 2a) to 38% reduction at 700°C (this process step is hereafter named as "DR to crossover", Section 3.3) under an $H_2$ flow rate of 0.5 L/min. In this work, the kinetically defined reduction degree of 38%, conducted at 700°C, was chosen as the "crossover point" for the hybrid reduction route (Section 3.3). Four 38%-partially reduced pellets obtained via DR (whose total weight was approximately 15 g) were transferred to a hydrogen reducing plasma (Ar-10%$H_2$) using the reactor schematically shown in Figure 2 (b) and processed according to the methodology described in Section 2.3 (this second step is hereafter named as "HPR after crossover"). The semi-reduced pellets were further reduced through HPR at different intermediate stages with exposure times to hydrogen plasma varying from 1 to 10 min. Supplementary Figure 3 shows the weight fractions of iron and wüstite for these samples, obtained using the same protocol described in Section 2.3. The corresponding weight fraction of wüstite was used to calculate the amount of oxygen in each sample.

## 2.5 Microstructural characterization

Representative samples processed via the hybrid reduction route (Section 2.4) were cut, embedded and metallographically prepared for high-resolution scanning electron microscopy analysis using a Zeiss Merlin scanning electron microscope (SEM). Energy-dispersive X-ray spectroscopy (EDS) analyses were also conducted for this sample at an accelerating voltage of 15 kV.

## 3. Results

### 3.1 Reduction kinetics of direct reduction versus hydrogen plasma reduction

Figure 3 (a) shows the solid-state DR kinetics of hematite pellets exposed to pure molecular hydrogen at 700 and 900 ºC under a constant gas flow of 0.5 L/min, as performed in a preceding study [30]. This figure also displays the corresponding kinetics for HPR of the molten iron ore, conducted with a small $H_2$ partial pressure of 10% (viz. Ar-10%$H_2$). Figure 3 (a) also reveals that the total conversion of hematite into liquid iron is achieved after 15 min of exposure to hydrogen plasma, whereas sponge iron is produced in the DR route within ~70 and ~40 min direct exposure of hematite to $H_2$ at 700 and 900°C, respectively.

The reduction rates for both DR and HPR are shown in Figure 3 (b). In this figure, the vertical dashed line at 33% reduction (i.e. 33 wt.% oxygen loss) corresponds to the theoretical reduction degree at which hematite is expected to be completely transformed into wüstite (through magnetite) during DR, i.e. 33% reduction represents the onset of the solid-state wüstite reduction regime [30]. However, hematite pellets, when exposed to a reducing gas, get heterogeneously reduced, with substantial gradients through their thickness [27,30,34]. This means that, at a 33% global reduction degree, iron forms preferentially at the outermost layers of the pellets or ore pieces, whereas their core remains partially reduced into magnetite and wüstite, as reported in a preceding study [30].

Figure 3 (b) also shows that the transformations beyond 33% reduction exhibit reaction rates of almost one order of magnitude lower than those observed at the earlier reduction stages, independently of the process employed. In the solid-state DR scenario, such sluggish kinetics is mainly associated with slow nucleation rates of iron domains within the oxide pellets as well as to sluggish mass transport of outbound oxygen through the outer dense iron layer of the pellets [30,31]. The use of HPR that simultaneously melts and reduces the ore, however, enables to proceed approximately two times faster from 33% reduction onwards, as shown in Figure 3 (b). This is because within the high-current plasma, high densities of highly energetic hydrogen-originated radicals are being created, including protons as the most reactive [13,17,35]. As they are accelerated towards the molten ore, their energy is partially transferred to the reaction interface, enabling local heat releasing and thus self-supplying energy to the reduction reaction [12,13].

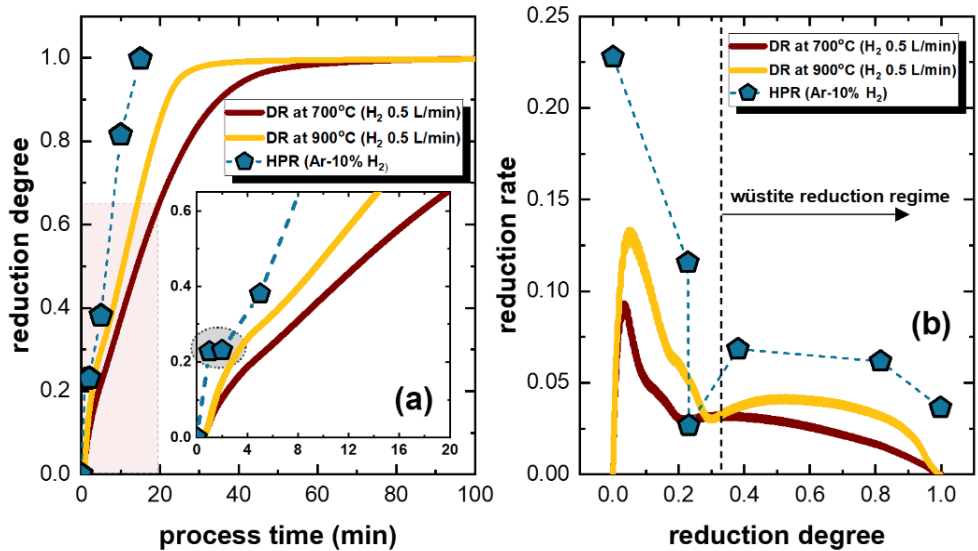

**Figure 3. (a)** Reduction kinetics of hematite pellets during hydrogen-based DR at 700 and 900°C ($H_2$ gas flow of 0.5 L/min), respectively, together with the corresponding reduction kinetics of molten iron ore performed via HPR using a lean Ar-10%$H_2$ gas mixture. **(b)** Reduction rates of both, DR and HPR as a function of the reduction degree. The vertical dashed line at 0.33 reduction represents the theoretical onset of the wüstite reduction regime during DR.

## 3.2 Process stability and thermal decomposition during hydrogen plasma reduction

The plasma arc stability was qualitatively evaluated at different stages of the HPR route, as shown in Figure 4 (a). At the early beginning of the process, especially when the iron ore is not yet completely melted, the arc is potentially unstable and gets scattered, as revealed in Figure 4 ($a_1$). Conversely, the arc becomes well-controlled and smoother when the material gets melted and the reduction proceeds further past the hematite to magnetite transition (Figure 4 $a_2$). This might be one of the possible reasons why reduction kinetics drastically vary at the beginning of the HPR process (i.e. in the early 20% reduction stage), as shown in Figures 3 (b) and (c). Figure 4 (a₃) shows an image of the specimen fully converted into a molten iron pool.

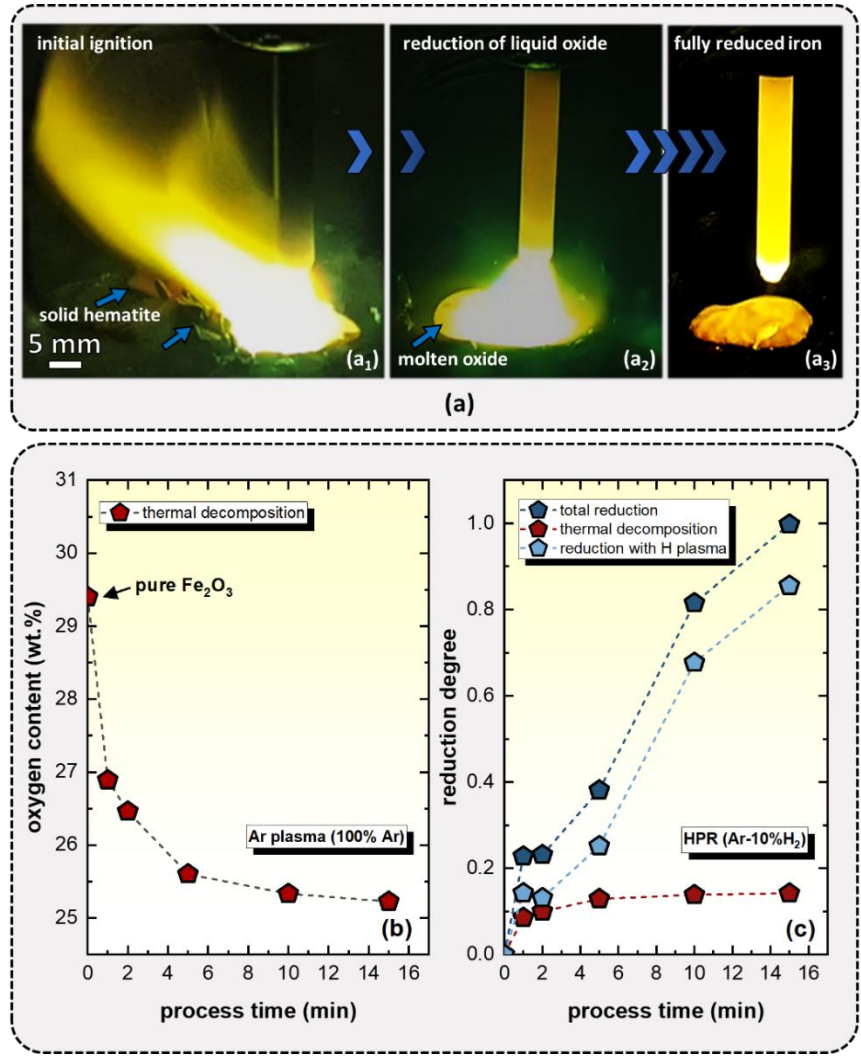

**Figure 4. (a)** Analysis of the arc plasma stability during HPR recorded at distinct reductions stages of the process. **(b)** Thermal decomposition of hematite during argon plasma processing. **(c)** Deconvolution of the competing partial chemical reactions during HPR: chemical changes due to thermal instability of hematite before melting, and the reduction associated with the redox reaction between oxygen of the ore and hydrogen plasma species.

Oxygen loss of hematite due to thermal decomposition was monitored under the same process conditions adopted in HPR, but using an argon atmosphere inside the arc-melting furnace rather than a gas mixture of Ar-10%$H_2$ (Section 2.3). The oxygen loss due to vapor pressure is reported in Figure 4 (b). This figure shows that over 15 min exposure to argon plasma, hematite reduces to a mixture of magnetite plus wüstite with a global oxygen content of approximately 25 wt.%, which translates into ~14 wt.% global reduction. These results evidence that the reduction degree achieved at each intermediate stage via HPR (as documented in Figure 3) is associated with two competing partial chemical reactions, namely, chemical changes linked to thermal instability of the ore before melting, and the reduction that is actually attributed to the redox reaction between oxygen of the ore and hydrogen plasma species, as reported in Figure 4 (c).

**3.3 Efficiency analysis**

The process efficiency of $H_2$ utilization during lab-scale DR and HPR was calculated by the ratio between the amount of $H_2$ effectively consumed by the reaction, which was derived from the kinetic data displayed in Figure 3 (a) for DR and Figure 4 (c) for HPR (i.e. after subtracting the thermal decomposition effects), and the hydrogen provided to the furnace (Section S.1 in the Supplementary Material). The efficiencies in $H_2$ utilization for both, DR and HPR follow the trends represented by the burgundy line and blue dashed line in Figure 5, respectively. Considering reduction solely via DR, the highest efficiencies are observed during the initial stages of the process, where hematite pellets become partially but quickly reduced as the hydrogen has surface access to the reaction front, as schematically shown in Figure 5. The efficiency of a full HPR route abruptly changes from 60% to nearly zero during the early 20% reduction stage (i.e. within the initial 2 min of the process where the arc tends to delocalize and where the reduction kinetics drastically vary, as shown in Figure 3). However, the efficiency increases again from 33% reduction onwards, displaying values between 20 and 40%. Figure 3 reveals that there is an optimal crossover point for these two technologies where their efficiencies in $H_2$ utilization are the same. These observations encouraged us to combine the best features of these two routes into a hybrid process to realize the most efficient use of hydrogen, as indicated by the green line in Figure 5.

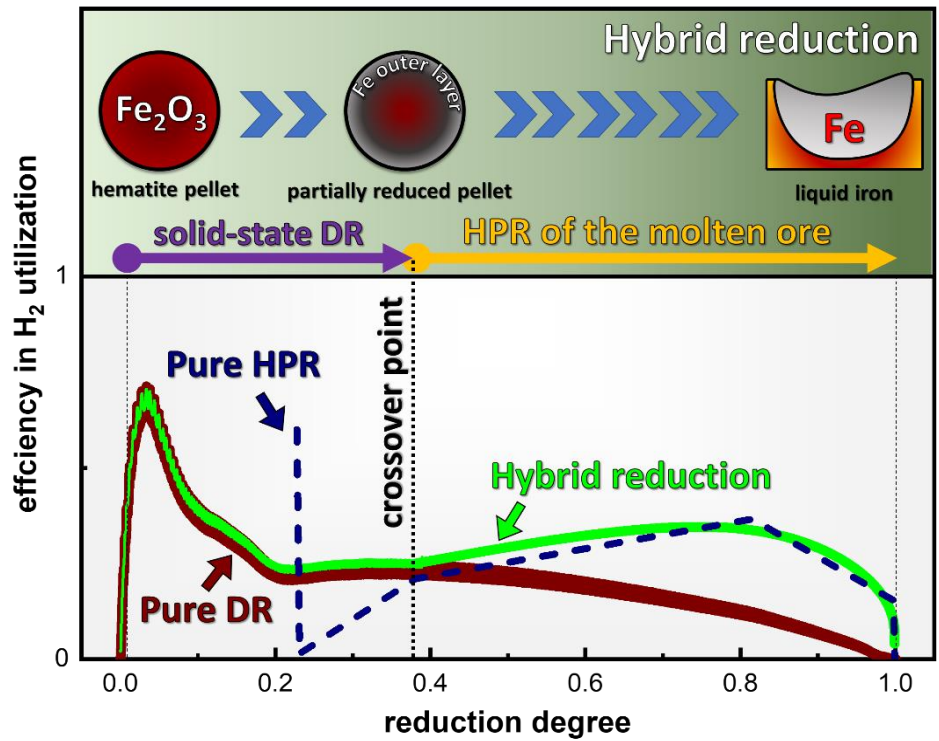

**Figure 5.** Overall trend of the efficiency in hydrogen consumption during a pure DR process (represented by the burgundy line), a pure HPR (blue dashed line) approach, and the corresponding one for a hybrid reduction (green line). For a hybrid scenario, hematite pellets are partially reduced to an optimal crossover point via DR, as schematically represented in the upper part of this figure. After the crossover point, the partially reduced pellets are transferred to an EAF with a reducing hydrogen plasma to complete the conversion into liquid iron.

Figure 6 (a) shows the efficiency plots for the pure DR, conducted at 700 and 900ºC, respectively, together with the corresponding one for the pure HPR. Opportunities for a crossover between the two processes occur at 38 and 60% reduction, respectively, when the direct reduction step is conducted at 700 and 900ºC. These results suggest that decreasing the temperature of the DR step allows its interruption at earlier reduction stages before transferring the material to the reducing plasma, thus enabling to reduce thermal energy consumption already in the DR step (it is worth to recall in that context that hydrogen-based direct reduction is endothermic).

The total amount of hydrogen circulated through the thermogravimetry system to produce sponge iron via a pure DR route is reported in Figure 6 (b) together with the corresponding $H_2$ quantities required in both steps of a hybrid process, namely "DR to crossover" and "HPR after crossover" (the latter using the arc-melting furnace). This figure shows that a hybrid process requires considerably less hydrogen (a total of 10.5 moles $H_2$/mol hematite to be flooded into the furnace) when compared with a pure lab-scale DR route performed at 700 or 900°C. However, the highest potential for saving hydrogen circulation is achieved by a hybrid process in which the "DR to crossover" step is conducted at 700°C up to 38% reduction. In this case, the amount of hydrogen employed to reach the crossover point is only 4.3 moles per mol of hematite. The subsequent "HPR after crossover" step requires only an additional amount of 6.1 moles $H_2$ /mol hematite, instead of 22.7 moles $H_2$ /mol hematite that would be necessary to fully convert the pellets into sponge iron at 700°C in the solely lab-scale DR route. Thus, in the hybrid scenario, six times less hydrogen is circulated through the system during the corresponding employed DR step. This observation also suggests that a hybrid procedure enables less amounts of hydrogen to be recycled from the off-gas stream before being reintroduced to the process (as reported in Supplementary Figure 2).

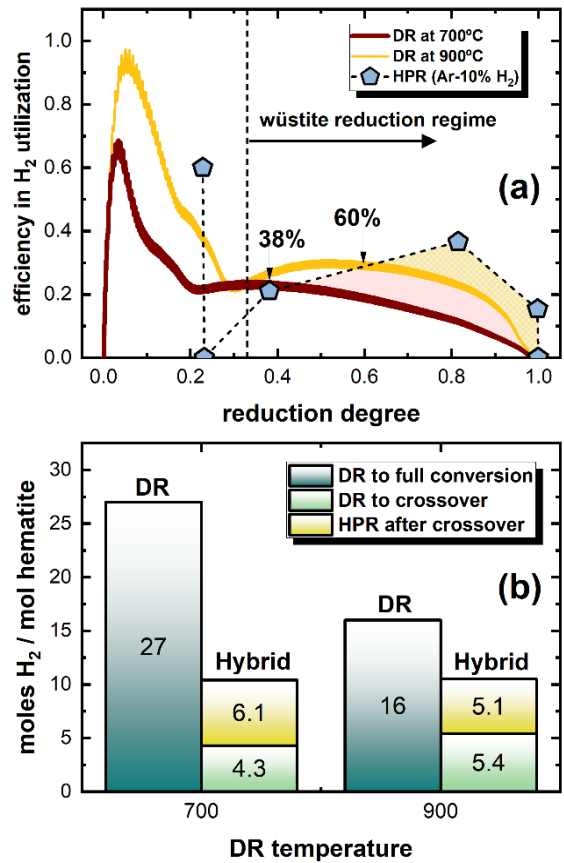

**Figure 6. (a)** Efficiency in H$_2$ consumption during DR of hematite pellets conducted at 700 (burgundy line) and 900°C (gold line) under a constant H$_2$ flow rate of 0.5 L/min. The corresponding H$_2$ efficiency during HPR is represented by the blue pentagon symbols. **(b)** Total amount of hydrogen employed to produce sponge iron via a pure DR route at 700 and 900°C, respectively. Corresponding hydrogen amount required in both subsequent steps of the hybrid processes, namely "DR to crossover" and "HPR after crossover".

**3.4 Validation of the hybrid reduction concept**

Based on the findings reported in Figure 4, we performed hybrid reduction experiments in which the "DR to crossover" step was conducted at 700°C to reach a reduction degree of 38%. Figure 7 (a) shows the reduction kinetics for the corresponding hybrid reduction route. The data reveal that both steps, namely "DR to crossover" and "HPR after crossover", require 10 min each. Figure 7 (b) reveals that the use of HPR after the crossover point enhances the reaction rates by about one order magnitude in comparison with the rates observed during the second half of the DR step (within the interval between 20 and 38% reduction). Figure 7 (c) summarises the total process time and the total amount of hydrogen employed in the hybrid reduction route proposed in this work. For the sake of comparison, this figure also shows the corresponding process time and $H_2$ quantities used in the stand-alone DR (conducted at 700 and 900°C) and HPR routes. Figure 7 (c) suggests that the hybrid reduction investigated here enables a total hydrogen consumption similar to that employed in a pure HPR scenario, which is considerably lower than those quantities required to produce sponge iron via DR.

Figure 7 (c) also shows that the total process time of the hybrid approach is 20 min, a value that is 5 min longer than the total time required to produce iron solely via pure HPR. However, Figure 7 (b) shows that the "HPR after crossover" step of the hybrid approach is only 10 min, and not 15 min as in the case of a pure HPR.

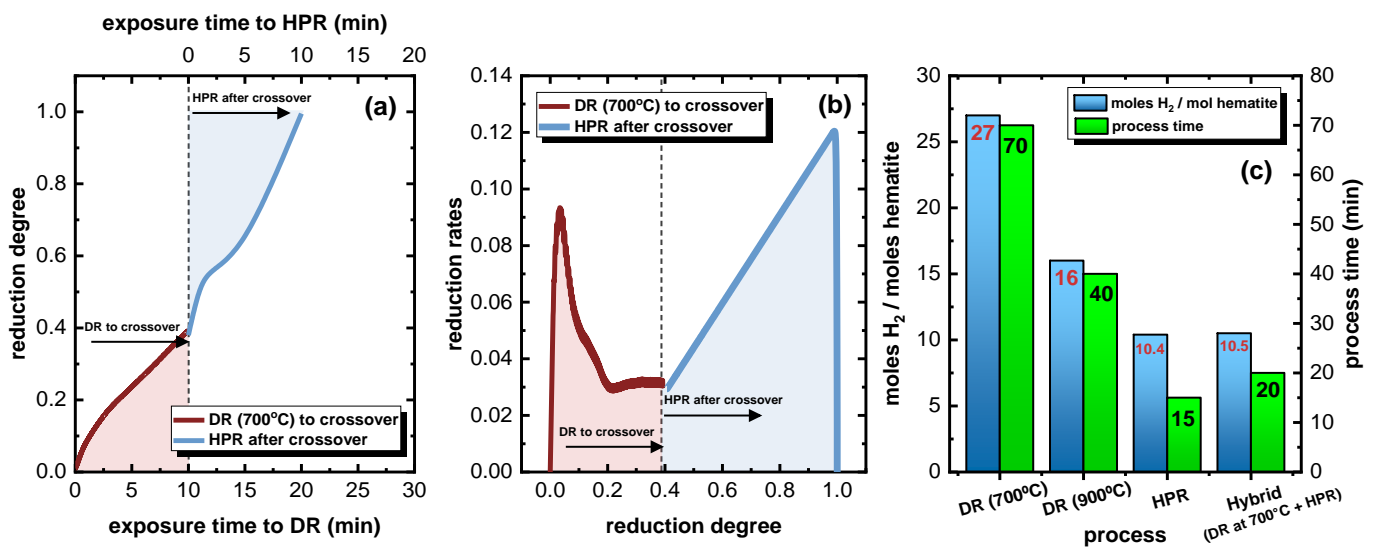

**Figure 7.** **(a)** Reduction kinetics of the hybrid route proposed in this work. The burgundy line represents the reduction interval conducted via DR at 700°C up to 38% reduction (crossover point), requiring 10 min. The blue line represents the reduction kinetics after the crossover point, which is performed under HPR. **(b)** Reduction rate of the hybrid route plotted as a function of the reduction degree. The burgundy line stands for the reaction rates conducted to the crossover point via DR. The blue line shows the reaction rates after the crossover point via HPR. **(c)** Comparison of the total process times and amounts of hydrogen consumed in the different routes for reducing iron ores, including the hybrid reduction route proposed in this work.

Figures 8 (a) and (b) display images of hematite subjected to different steps of a hybrid reduction process. Figure 8 (a) shows one hematite pellet before and after partial direct reduction to 38% at 700°C. Four 38%-partially reduced pellets were transferred to HPR and processed for 1 and 10 min, achieving ~ 50 and ~ 100% reduction, respectively (Figure 8b). After solidification, these specimens were probed via microscopy at the regions indicated by the numbers 1 to 4 in Figure 8 (b). The major portion of iron produced after 1 min of HPR sinks to the bottom of the melt pool (region of the sample with bright contrast in Figure 8b), leaving the remaining unreduced oxide on the top of the sample [12,18]. Microstructural characterization of this sample (Figure 8c), reveals that small iron dendrites are found dispersed within the remaining wüstite (frames 1 and 2 in Figure 8e), which solidifies with a typical dendritic structure. Iron obtained after 10 min of HPR contains a small volume fraction (~ 0.01) of Fe- and Si-bearing oxide particles that remained trapped within the solidified material, as revealed by the elemental distribution maps displayed in Figure 8 (d). These particles have also a tendency of floating atop of the melt pool (similar to a slag layer), as indicated by the arrow in Figure 8 (b), due to their lower density when compared with that of iron. Therefore, it is conceivable that they can be properly separated from the melt when adequate conditions for flotation are applied.

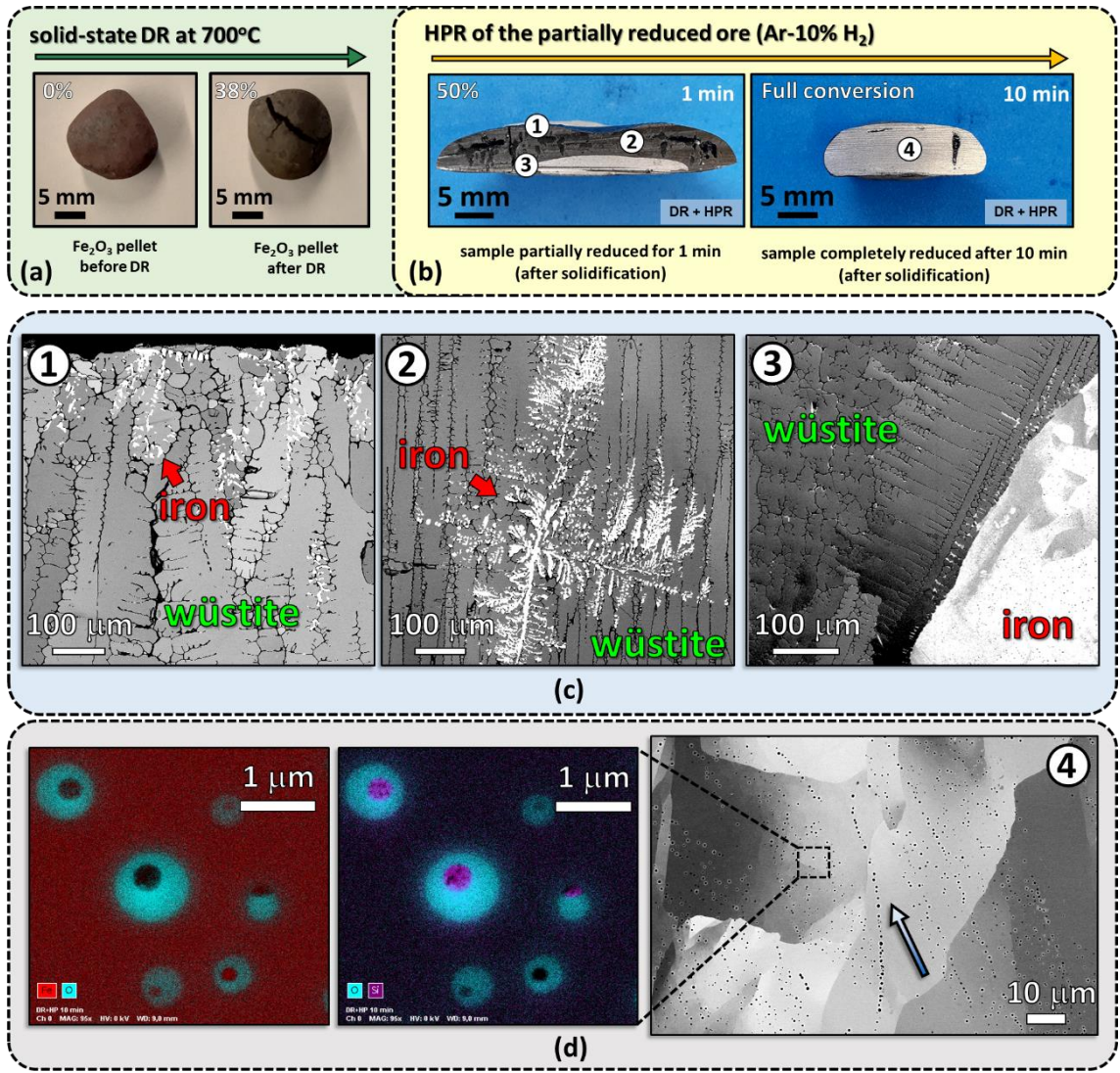

**Figure 8.** **(a)** Hematite pellet before and after 38% partial reduction conducted via DR at 700°C under an $H_2$ flow rate of 0.5 L/min. **(b)** Four partially reduced pellets were subjected to subsequent HPR and further processed for 1 and 10 min, achieving total reduction degrees of ~ 50 and ~ 100%, respectively. **(c)** Microstructural characterisation of the specimen partially reduced to 50% reduction (1 min of HPR) conducted at the regions labelled as "1", "2" and "3" in (c). **(d)** Microstructural characterisation of the final iron product. Approximately 1% of the material is composed of unreduced Fe- and Si-containing oxide particles, as revealed by the elemental distribution maps. These particles tend to float in the melt, as indicated by the blue arrow in the backscatter electron image.

## 4. Discussion

Hydrogen-based iron production is the most viable option to drastically reduce the staggering $CO_2$ emissions in this sector. In that context, the economical consumption of this reductant is of the highest importance for sustainable iron production, as green hydrogen produced by renewable energy sources will continue as one of the major bottleneck during the next decades [23,36,37]. The underlying reasons are the low electrolysis productivity and the insufficient capacities in both renewable electricity and industrial infrastructures to produce sufficient amounts of green hydrogen, especially in view of the gigantic demand for currently 1.8 billion tons of steel being produced every year, with forecasts predicting 2.4 billion tons by the year 2040 [38]. In its current transient technology state, hydrogen comes mainly from reforming methane via steam (called grey hydrogen, or blue when it is combined with carbon capture and storage [23]). Therefore, highly efficient exploitation of hydrogen for carbon-free iron-making is crucial for the challenge of decarbonizing the steel industry [1], enabling also lesser efforts for hydrogen recycling and reutilization.

The pure HPR route for green iron production would be an ideal scenario, where the ore is simultaneously melted and reduced. However, our results obtained in model laboratory-scale processes suggest that significant technical adjustments for plasma arc stabilization at the beginning of the process (Figure 4 $a_1$) are necessary in order to reduce process inefficiency with respect to reduction kinetics (Figure 3), electrical energy consumption and use of hydrogen (Figure 6). An important processing aspect is that the initial arc ignition is rather violent and unstable, causing sputtering of solid hematite which delays the progress of the reaction and can damage the electrodes. Prolonged exposure to an unstable arc and violent bath dynamics can also cause damage to the inner structure of the reactors (e.g. the refractory linings) when not protected by purposely developed slags, diminishing their lifespan and requiring additional maintenance costs [32]. Conversely, the plasma is much smoother and well-controlled when the reaction proceeds from 33% reduction onwards, permitting better exploitation of hydrogen, as documented in Figure 6 (a). These factors concerning arc stability, slag development and efficient use of hydrogen are essential aspects for up-scaling of the HPR process.

Considering that DR starts efficiently and fast during the first stages of the reduction and that sponge iron currently produced via industrial DR routes must be melted anyway in EAFs, a hybrid strategy involving DR and HPR allows to combine the advantages of both technologies: first reducing hematite to a kinetically defined reduction level with controlled and efficient DR, to later transfer the semi-reduced product to a plasma furnace enabling the enhancement of both kinetics and $H_2$ utilization at simultaneously improved process stability. Our research results for the hybrid route provide promising insights on efficiency and kinetics in this direction. The total hybrid reduction time is only 5 min longer than that required in a pure hydrogen plasma reduction (Figure 7). However, we must consider that the exposure time to plasma in the hybrid route is shorter (i.e., only 10 min and not 15 min as in the case of a solely HPR, Figures 3 and 7), enabling a longer lifespan of the reactor linings and electrode materials. The total amount of hydrogen provided in the hybrid process is practically the same as the one employed in an ideal pure HPR scenario, i.e. the approach of bringing partially reduced pellets or lump ores into a reducing plasma environment produces liquid iron at minimal hydrogen utilization yet enabling optimized opportunities for its recycling (Supplementary Figure 2). Recycling of $H_2$ is another fundamental aspect in existing hydrogen-based steel production, as it allows not only recovering costly $H_2$ but also the associated thermal energy [15,39]. As reduction of iron ores conducted solely via hydrogen-based DR is endothermic, an excess of hydrogen input is demanded to serve as heat carrier in shaft-furnaces and maintain the productivity rates [40]. Typical numbers associated with the MIDREX technology and other reports [41] are ~800 $m^3$ of $H_2$ per ton of produced sponge iron, of which 250 m$^3$ are essentially required as fuel for gas heater. Energy recovery from H$_2$ in condensers are dependent on the excess of hydrogen flooded in shaft-furnaces though, as recently documented in a detailed assessment of process parameters for industrial hydrogen-based reduction of iron ores. It was demonstrated that an oversupply of hydrogen above six times the quantity required to produce sponge iron hinders heat recovery from hydrogen output in condensers [15]. This means that additional energy supply must be given to the system to heat the input hydrogen. Results from our lab-sale hybrid route suggest that lower quantities of H$_2$ could be employed in the DR step, as the product is not sponge iron but a semi-reduced oxide processed at relatively low temperatures. In other words, DR is only conducted within the hematite through magnetite to wüstite reduction regime, avoiding entering the wüstite reduction interval that thermodynamically requires higher chemical potentials of hydrogen to drive the kinetically sluggish reaction FeO + H$_2$ → Fe + H$_2$O. This implies that the residence time for the input material in a shaft furnace can be shortened, yet suggesting that reactors with more compact dimensions could be designed for this purpose. When processing the semi-reduced material during the second step of the hybrid route, namely HPR, an efficient usage of H$_2$ is also indispensable, as the off-gas is not only composed of non-consumed H$_2$ and product water, but it also contains a broad variety of gangue elements (e.g. P, S, Si and others) that get evaporated from the ore due to their high vapor pressures [12]. Thus, additional efforts are demanded for H$_2$ purification before being reintroduced to the process if a non-optimized usage of H$_2$ is conducted during the HPR step.

Although our results are based on laboratory experiments, and the absolute values reported here might change when upscaled to large-scale industrial aggregates, the results suggest that a viable hybrid strategy for the green steel industry can be applied, based on merging the best features of DR and HPR at minimal changes to existing industry-scale aggregates and workflows, with special emphasis on the efficient use of energy and hydrogen, both key bottlenecks in green steel making.

## 5. Summary and conclusions

We investigated the fundamental and process aspects of a hybrid hydrogen-based reduction of iron ores. Based on lab-scale experiments, the maximum utilization and highest potential for $H_2$ savings were achieved when hematite pellets were partially reduced to 38% at 700°C with a $H_2$ flow (namely, crossing over point), and subsequently converted into liquid iron under a hydrogen reducing plasma (Ar-10%$H_2$).

Results also suggested that this hybrid route allows to exploit the best kinetics reduction, energy and stability aspects in both DR and HPR steps. This is because on one hand the solid-state phase transformations that heterogeneously occur along the pellets length until reaching 38% global reduction (i.e. before the crossing over point) proceeds fast and efficient. On the other hand, the arc plasma proceeds much more stable when processing semi-reduced oxides (e.g. 38%-reduced pellets) in comparison with the scenario where hematite is used as input material. Our findings suggested that six times less hydrogen is consumed in the DR step of a hybrid process when compared with a pure lab-scale DR route conducted at 700°C. This suggests that the residence time in direct reduction furnaces can be shortened, as the product is not sponge iron but a semi-reduced oxide. Therefore, reactor with more compact dimensions could be designed for this purpose. These findings also suggest that less costly $H_2$ can be sent to purification before being reintroduced to the process, a task that would require lesser efforts, energy and costs. Better utilization and recycling during HPR are also of great importance as non-consumed $H_2$ present in the off-gas is mixed with water vapor, argon, and gangue elements evaporated from the ore, thus requiring additional efforts to its purification.

With this work we aim to provide a decision-making suggestion for sustainable ironmaking in order to avoid energy expenses for unnecessary recirculation of $H_2$ in shaft furnaces or plasma rectors, yet permitting the production of pure iron with competitive conversion rates.

**Acknowledgements**

We thank Monika Nellessen and Katja Angenendt for their support to the metallography lab and SEM facilities at MPIE. We are grateful to Benjamin Breitbach for the support to the X-ray diffraction facilities at MPIE. We appreciate the support to the thermogravimetry apparatus at MPIE provided by Dr. Dirk Vogel and PD Dr. Michael Rohwerder. IRSF acknowledges financial support through CAPES (Coordenação de Aperfeiçoamento de Pessoal de Nível Superior) & Alexander von Humboldt Foundation (grant number 88881.512949/2020-01). HS acknowledges the financial support through the Heisenberg Programm of the Deutsche Forschungsgemeinschaft (grant number SP1666/1-2) YM acknowledges financial support through Walter Benjamin Programm of the Deutsche Forschungsgemeinschaft (project number 468209039). AM acknowledges financial support through Alexander von Humboldt Foundation (Humboldt-ID: 1215046).

**Author Contributions**

IRSF was the lead scientist of the study; ISRF, HS and DR designed the research and wrote the paper; ISRF and HS performed the calculations and evaluated the data. AM, YM, CCS conducted the direct reduction experiments. MK conducted the hydrogen plasma reduction experiments. All authors discussed the results and commented on the manuscript.

**Competing Interests**

The authors declare no competing interests.